\documentstyle[12pt,psfig]{article}
\newcommand{\br}[1]{\mbox{$| #1 \rangle $}}

\begin{document}

\title{Quantum Computer Using Coupled Quantum Dot Molecules}
\author{Nan-Jian Wu\footnote{Electronic mail: nanjian@montblanc.ee.uec.ac.jp}, M. Kamada, A. Natori and H. Yasunaga \\
{\small \it{University of Electro-Communications, Chofu, Tokyo 182-8585, Japan}}}
\date{\small{9 December 1999}}
\maketitle


\begin{abstract}
We propose a method for implementation of a quantum computer using artificial molecules. The artificial molecule consists of two coupled quantum dots stacked along z direction and one single electron. One-qubit and two-qubit gates are constructed by one molecule and two coupled molecules, respectively. The ground state and the first excited state of the molecule are used to encode the \br{0} and \br{1} states of a qubit. The qubit is manipulated by a resonant electromagnetic wave that is applied directly to the qubit through a microstrip line. The coupling between two qubits in a quantum controlled NOT gate is switched on (off) by floating (grounding) the metal film electrodes. We study the operations of the  gates by using a box-shaped quantum dot model and numerically solving a time-dependent Schridinger equation,  and demonstrate that the quantum gates can perform the quantum computation. The operating speed of the gates is about one operation per 4ps. The reading operation of the output of the quantum computer can be performed by detecting the polarization of the qubits.

\end{abstract}

\section{Introduction}

One of the challenges in nanoelectronics is to develop quantum  computer that is based on solid-state devices.Any performance of quantum  computer can be decomposed into  a succession of operations of basic quantum gates: one-qubit gate and  two-qubit  controlled  NOT (CN) gate. To construct  a quantum computer, we must create  feasible basic quantum gates. This paper proposes a novel method for implementation of quantum gates using coupled quantum  dot molecules. 
Some implementations of the quantum gates have been proposed using solid-state structures. The main proposed gates are based on the dynamics of electrons in quantum dots$^{1,2}$,donor ions in Si$^3$ and Josephson junctions$^{4,5}$.Although the numerical studies of the quantum dot gates and the experimental demonstration of the Josephson junction qubit have been made, but it is not yet clear whether these implementations can be extended to many-bit system and large-scale quantum computer.$^{1,2,4,5}$  Secondly, two basic states \br{0} and \br{1} of the qubit are established by applying voltage or magnetic field so that the coupling of the electrons with external degrees of freedom spoils the unitary structure of quantum evolution (if the external voltage and magnetic field fluctuate). Furthermore,  the obvious obstacle to building the Si-based quantum computer$^{3}$ is the incorporation of the donor array into the Si substrate.  The quantum computer requires that the donor  must be placed  precisely into each qubit. It is extremely difficult to create the donor array by existing technologies, such as ion implantation and focused deposition. 
 
In the paper we propose a novel method for implementation of quantum gates that use artificial molecules. The artificial molecule consists of two coupled asymmetric quantum dots stacked along z direction. One-qubit and two-qubit gates are constructed by one molecule and two coupled molecules, respectively.The method for implementation of the quantum computer has the following features. 1) The structures of the basic  quantum gates are simple and can be  fabricated by existing technologies. 2) The implementation can be extended to a large-scale quantum computer.  3) Except for the signals that  control the quantum computation process, external electric field and magnetic field are not necessary to be applied to the gates so that the coupling of the electrons with external degrees of freedom can be reduced. In the following sections  we first present the structures of quantum gates and describe their operations (\S 2).  We then  establish analytic models of  the quantum gates and study numerically the operations and decoherence of the quantum computer (\S 3).  We also discuss issues related to the physical implementation (\S 4). Finally, we will summarize the main results (\S 5).

\section{Quantum Logic Gates}

\subsection{Qubit}

Figure 1 shows the schematic structure of the proposed qubit using an artificial molecule that consists of two stacked asymmetric quantum dots with a disk shape and one single electron. The single electron can tunnel between two quantum dots. The dimensions  of the asymmetric quantum dots are  designed well so that the ground  and first excited sates are mainly localized in the dot 1 and dot 2, respectively, as shown in Fig.1(c). The two localized states correspond to the bonding and anti-bonding states of the coupled quantum dots, respectively. The energy states of the artificial molecule can be used to represent \br{0} and \br{1} states of the qubit.  
	
The two states of the qubit exhibit two bistable polarizations of the electron charge and  are manipulated by an electromagnetic (optical) resonance or voltage pulse$^{5-7}$. Here we use a resonant electromagnetic wave  to irradiate and manipulate the qubit through a microstrip line integrated on the substrate.The approach has the following advantages: 1)A qubit can be addressed by a microstrip line connected directly to it; 2)it can be realized by the existing nano-fabrication and large scale integrated circuit (LSI) technologies; and 3) comparing with other kinds of qubits in which the \br{0} and \br{1} states are prepared by biasing external electrical field or magnetic field,  the coupling of the electrons in our qubit with external degrees of freedom can be reduced. A role of the metal films in the qubit structure will be explained later. 

\subsection{Two-Qubit Controlled NOT(CN) Gate}
We can constitute a quantum CN gate using two qubits with different resonance frequencies. Figure 2 shows the schematic structure of the CN gate that consists a control qubit and a target qubit. The diameters of the quantum dots in the control qubit  are designed to be different from  the diameters of the dots in the target qubit, respectively. The coupling between the two qubits are controlled by three metal film electrodes between the qubits: upper electrode, middle electrode and lower electrode.  The middle electrode is always grounded.  We numerically calculated the electrostatic  interaction between the electrons in the qubits. If the upper and lower electrodes are  floated, the electrons at two kinds of  configurations: 1) occupying  dot 1 (dot2) and dot 3 (dot4) and 2) occupying dot 1 (dot2) and dot 4 (dot3) can be coupled by Coulomb repulsion interactions U and U$_1$,respectively. But, Coulomb repulsion interaction U is much larger than U1.Therefore  the two qubits are coupled by the dipole-dipole interactions. The electronic state of the target qubit depends on the state of the control qubit. On the other hand,  if  all of the electrodes  between the qubits are grounded, it is found that  the interaction energy between the electrons can be controlled to be smaller than 10$^{-10}$ eV.  Consequently,  Coulomb interaction  between the qubits can be  turned off, that is, the energy levels of the two-qubit gate can  be controlled by turning on/off the electrostatic interaction between two electrons, as  shown in Fig. 2(c). 

The quantum CN gate operates as follows. First we turn on the Coulomb interaction between two qubits by floating the upper and lower electrodes. Then we use the resonant electromagnetic wave of the frequency $\omega_{T+U}$ to irradiate target qubit for $\pi$-pulse time. If the control qubit is set to \br{0}, the target qubit does not change its state, and if the control qubit is \br{1}, the target qubit change its state from  \br{0}(\br{1}) to \br{1}(\br{0}). Therefore, the XOR operation is reached. Finally the Coulomb interaction is turned off,  the state of one qubit becomes independent on that of the other qubit.  The CN gate has the advantages: 1)  the coupling between the qubits can simply controlled by floating or grounding  the metal electrodes;   and 2)  even if the length of the metal electrode between two qubits  is made longer, the Coulomb repulsion interactions  U  can be maintained  so that the array of the qubits in the quantum computer is designed easily.

\section{Operation of Quantum Computer}
We will analyze numerically the operation of the quantum logic gates by the method described below and demonstrate that the proposed quantum gates can perform quantum computation.   
\subsection{Model}
The qubit consists of two coupled asymmetric quantum dots(1 and 2)stacked along the z direction.  The size of the dot 1 is set to be  larger than the dot 2.  We use a  box-shaped potential as an analytic model of the quantum dot.  The valence-band and core electrons are neglected in  our analysis because these electrons are highly localized and their wave functions do not overlap with each other.  The dimensions  of the asymmetric quantum dots are designed well  so that the ground and first excited sates of the coupled quantum dots are mainly localized in the dot 1 and dot 2, respectively.We use the basic model to investigate  the eigenstates and the quantum dynamics of the quantum computation logic gates.The quantum logic gates can be described by the following Hamiltonian

\begin{eqnarray}
H &=& \mathop{\sum}_{i=1,2}H_i+\mathop{\sum}_{k,j=1,2}U_{1k,2j}+\mathop{\sum}_{i,j=1,2}Qz_{ij}E_{zi}COS(\omega_i t) \\
H_i &=& \mathop{\sum}_j \left\{ -\frac{\hbar^2}{2m} \nabla_{ij}^2+V_{ij}(x,y,z) \right \}\\
V_{ij} &=& \left\{
\begin{array}{ll}
0 & x_{ij0}-\frac{w}{2} \leq x \leq x_{ij0}+\frac{w}{2},y_{ij0}-\frac{w}{2} \leq y \leq y_{ij0}+\frac{w}{2} \\
V_0 &  x<x_{ij0}-\frac{w}{2},x>x_{ij0}+\frac{w}{2},y<y_{ij0}-\frac{w}{2},y>y_{ij0}+\frac{w}{2} \\
V_1 & z<z_{ij0}-\frac{h}{2},z>z_{ij0}+\frac{h}{2}
\end{array}
\right.
\end{eqnarray}
where $H_i$ represents the Hamiltonian of the i-th qubit with an electron effective mass m ,$V_{ij}(x,y,z)$ is the box-shaped three-dimensional confinement potential giving rise to  the j-th quantum dot in the i-th qubit,  $U_{1k,2,j}$ describes the Coulomb interaction between the electrons occupying dots k and j in the two qubits 1 and 2.The interaction between the resonant electromagnetic wave of the amplitude $E_{zi}$ (along z direction) and the frequency i and the electron in the j-th dot of the i-th qubit is described by $Qz_{ij}E_{zi}cos(\omega_{i}t)$. $Q$ is electron charge. w and h are width (x,y) and height of the dot, respectively. $(x_{ij0}, y_{ij0}, z_{ij0})$ are the center coordinates of the j-th dot in the i-th qubit. We first obtain numerically the eigenstates  of one qubit.Then we construct the basic functions:  the two basis vectors of \br{0} and \br{1} for one qubit, and the four basis vectors of \br{00},\br{01},\br{10} and \br{11} for the CN gate. After that we solve a time-dependent Schridinger equation numerically and analyze the operation of the quantum logic gates. 

Finally we analyze the electron-phonon scattering in one qubit and estimate the decoherence time by taking electron-phonon interaction in the coupled quantum dots at low temperature ( about 0 K). For sake of simplicity, we assume the energy difference between the \br{0} and \br{1} states is smaller than the optical-phonon energy.  Therefore longitudinal-acoustic (LA) phonon emission (adsorption) via deformation potential interaction is considered. The deformation potential energy is given by$^8$

\begin{equation}
V_{e-ph}(r,q,t)=i \sqrt{\frac{\hbar q}{2\Omega \rho U_s}}D[e^{iqr}e^{-i\omega_q t}+e^{-iqr}e^{i\omega_q t}].
\end{equation}
where $\Omega$ is the system volume, D is the deformation potential constant,is the mass density, q is wave vector and Us is sound velocity. We assume a linear dispersion relation, $q = qU_s$.The confinement effect of the phonon modes is ignored, that is known to be legitimate for LA phonon.$^{12}$
	
\subsection{Operation of Quantum Logic Gates}
   The above model is applied to a concrete example of the qubit that is  formed by the stacked GaAs/AlGaAs quantum dots. Figure 3 shows the calculated results of the operation of the qubit. The dependence of the probabilities $|\alpha|^2$ and $|\beta|^2$ of the \br{0} and \br{1} states on the irradiation of the electromagnetic wave obtained.  The dielectric constant was taken to be 10. The effective mass of the electron is 0.67m$_0$.The electron confinement potential $V_0$ and $V_1$ are 1eV and 0.24eV, respectively.The energy difference between the \br{0} and \br{1} states of the qubit was set to about 25 meV.  The frequency and the amplitude (electric field) of the resonant electromagnetic field is 6 THz and 1.5mV/nm, respectively.  As shown in Fig. 3, the qubit changes its states from \br{0} (\br{1}) to \br{1} ( \br{0}) due to irradiation of the electromagnetic wave for the $\pi$-pulse time.The operating speed of the gate is very fast(about operation per 4ps). On the other hand, when a $\pi /2$-pulse of the electromagnetic wave was irradiated on the qubit, the operation of Hadamard transformation

\begin{equation}
\br{0}\rightarrow \frac{1}{\sqrt{2}}(\br{0}+\br{1}),\:\br{1}\rightarrow \frac{1}{\sqrt{2}}(\br{0}-\br{1}).
\end{equation}
was given. 

Figure 4 shows the calculated results of the operation of the two-qubit CN gate. The resonant energies of the control and the target qubits were set to 26 meV and 18 meV, respectively. The Coulomb repulsion interaction energy U was taken to 10meV. As shown in Fig. 4, it was confirmed that if the control and target qubits are set to different states, respectively, such as \br{0},\br{1} and $\alpha \br{0}+\beta \br{1}$, the XOR operation and the entanglement states were realized by applying intelligently the resonant electromagnetic wave with frequency $\omega_{T+U}$. The above results demonstrated that the proposed gates can be used to construct the quantum computer and perform quantum computation.

Considering the electron-phonon interaction in a qubit,  we calculated the dynamics of the electron in the qubit. Figure 5 gives the time evolution of the state of the qubit with the phonon-electron interaction. The energy difference between the \br{1} and \br{0} states was set to 26 meV, and the phonon energy was taken to satisfy the energy conservation. The deformation potential D is -6.8 eV, the density  is 5.4$\times 10^3$ kg/m$^{3}$, and the sound velocity $U_s$ is 3.4$\times$ 10$^{3}$ m/s.The figure shows the evolution of the probability of the \br{0} state of  the qubit with the electron-phonon interaction under the irradiation of the electromagnetic wave.To compare with the result without the electron-phonon interaction, the Rabi oscillations of the qubit under the irradiation of the electromagnetic wave is given also.Form the results, it was found that due to the phonon-electron interaction the obvious phase  shift of Rabi oscillation was observed after 10 ns. The electron-phonon scattering rate was estimated to be about $\sim10^7$/s 

\section{Discussion}

We will discuss several subjects related to the fabrication, operation and the reading.The rapid progress in the nano-technology made it possible to fabricate fine quantum dot structure with small size of 5$\sim$100nm.Furthermore, several  techniques, such as mesa etching of triple barrier heterostructure$^{6,7}$,and self-organization growth$^9$, have been proposed to produce a two-dimensional array of the stacked asymmetric quantum dots. The qubit in our quantum computation can be implemented by combining the above techniques and the standard semiconductor process used in LSI fabrication. Indeed,some kinds of the quantum artificial molecules have been fabricated by several groups.$^{6,7,10}$ After a two-dimensional array of the quantum artificial molecules are formed,the metal film electrodes and the insulator films in our quantum computation can be  produced by the film deposition,lithography and etching process. 

We address and operate one qubit in a quantum computer by selectively supplying an  electromagnetic power to a chosen microstrip line.When a qubit is biased by the electromagnetic power, the microstrip lines that connected to the other qubits are  grounded so that the other qubits are screened from the electromagnetic wave. We estimated the structure of the microstrip line needed to the qubit.  Assuming that the resonant energy of the qubit  ranges form 10 to 100 meV,  the frequency  of the resonant electromagnetic field is between 2.4 and 24THz.  The progress in the microstrip line technique is rapid.$^{11}$ If we design both of the height and width of microstrip line are about 1 $\mu m$ and the separation between the line and the substrate electrode is about 1 $\mu m$, the microstrip line can be integrated  by LSI process, on the semiconductor substrate.  Figure 6 gives a schematic layout of the quantum computer. 

In our qubit there are many physical mechanisms contributing quantum decoherence, such as phonon scattering, spontaneous emission, impurity scattering and interface trap scattering. But,  the spontaneous emission results in a lifetime longer than that by the phonon-electron interaction.$^{12}$ Although the impurity scattering result in  departures from unitary structure of quantum evolution and the lifetime of 10$^{-9}$s,$^1$ in principle,impurities can be reduced by  improving the crystal growth and nano-fabrication technologies. The coupling of the electrons to the phonons is unavoidable.We estimated that the relaxation time of the electron-LA phonon scattering is about 10$^{-7}$s and that a raito of the relaxation time to operating time of  gate is about 10$^{4}$ $\sim$ 10$^{5}$ at T = 0K.  At finite temperature, the electron-LA phonon scattering rate can be obtained by multiplying the zero-temperature result by $(1+n_q)$ (due to the simulated emission).$n_q$ is LA phonon (Bose-Einstein) distribution function.$n_q=1/[exp(h \omega_q/KT)^{-1}]$,where T is the phonon temperature, K is Boltzmann constant and $h \omega_q$ is the energy of the phonon with wave vector q. To satisfy the energy conservation, $h \omega_q$ is taken to be about 20meV in the quantum gate. When the temperature is below T = 77K,  $1+n_q$.This means that the designed qubit and CN gates can work below T = 77K.  Furthermore, if we design a qubit so that the energy difference between the \br{0} and \br{1} is much larger than of the energies of optical and acoustic phonon, the electron-phonon scattering  rate may be lowered by the well-known phonon bottleneck phenomenon,  and the maximum possible operation temperature can be  raised also.

The reading operation of the qubit can be performed by detecting the voltage potential. We have calculated the voltage potential  near the upper dot of the qubit  and discovered that the voltage potential distribution changes markedly when the state of the qubit is switched. Figure 7 shows typical contour plots of voltage potential distributions on the cross section of the qubit.If we switch  the \br{0} and \br{1} states of the qubit, the potential difference at measuring point A can reach 4.6mV. After the quantum computation is finished, we can use electrometers to measure the potential difference through the microstrip lines and to read out the output of the quantum computer. Details of the operations will be investigated in the future. 

\section{Summary}
We proposed a method for implementation of a quantum computer using artificial molecules. The artificial molecule consists of two coupled quantum dots stacked along z direction and one single electron. The ground state and the first excited state of the molecule represent the \br{0} and \br{1} states of a qubit, respectively.The state of the qubit is manipulated by a resonant electromagnetic wave that is applied directly to the qubit through a microstrip line. The coupling between two qubits in a quantum controlled NOT gate is switched on (off) by floating (grounding) the metal film electrodes. We studied numerically the operations of the  gates and demonstrated that the quantum gates can perform the quantum computation. The operating speed of the gates is about operation/4ps. The estimated the decoherence time is about 10$^{-7}$s.The reading operation of the output of the quantum computer can be performed by using electrometers to detect the polarization of the qubits.

\newpage

\section*{Figure Captions}
Figure 1. Schematic structure of a qubit using two coupled quantum dots (an artificial molecule): (a) a top view of the qubit; (b) a  cross section of the qubit; (c) electron potential along z direction and the two lowest energy levels of the qubit. The structure of the qubit contains metal film electrodes and a microstrip line. 
\\ \\
Figure 2. Schematic structure and energy levels of  a two-qubit controlled NOT (CN) gate: (a) a top view of the CN gate; (b) a cross section of the CN gate; (c) energy levels of the gates when the coupling between the qubits is turned  on or off.   If the coupling between the qubits is turned off, T  and C are the resonant frequencies of the electromagnetic waves; if the coupling is turn on, $\omega_{T+U(C+U)}$ and $\omega_{T-U(C-U)}$ are the resonance frequencies of the target (control) qubit, respectively, when the control (target) qubit is set to \br{1} and \br{0}. Here Coulomb interaction between the electrons at dot 2 (1) and 4 (3) is ignored.  
\\ \\
Figure 3. Calculated basic and Hadamard transformation operations of the qubit. The probabilities of the \br{0} and \br{1} states of the qubit $\br{\psi}=\alpha \br{0}+\beta \br{1}$ were given. $\omega_r$ is the resonance frequency of the qubit. $\pi$ and $\pi /2$ represent the $\pi$ and $\pi /2$ pulse operations, respectively. (Dot 1 : w = 24nm and h = 20nm; dot 2: w = 22 and h = 15nm; separation between two dots is 7nm)
\\ \\
Figure 4. Calculated basic operations of the CN gate and an example entanglement state. The probabilities of the \br{00}, \br{01}, \br{01} and \br{11} states of the gate ($\br{\psi}=\alpha \br{00}+\beta \br{01}+\gamma \br{10}+\eta \br{11}$) were given. $\omega_{T+U}$ is the resonance frequency of the target qubit.  $\omega_{rC}$ is the resonance frequency when the coupling between the two qubit is turned off. $\pi$ and $\pi /2$ represent the $\pi$ and $\pi /2$ pulse operations, respectively. (dimensions of the control qubit is the same as that in Fig.3;  dot 3  : w = 29nm and h = 20nm; dot 4: w = 27 and h = 15nm; separation between two dots is 7nm)
\\ \\
Figure 5. Time evolution behavior of the state \br{0}  of the qubit with the phonon-electron interaction at near 0ns, 10ns and 50ns. To compare the result of the qubit without  phonon-electron interaction, Rabi oscillations of the qubit under the irradiation of the electromagnetic wave is given also. 
\\ \\
Figure 6. A schematic structure of  the quantum computer based on the artificial molecules.  If a metal  partition is deposited between the microstrip lines, the influence of the electromagnetic wave on the neighbor qubit can be removed.\\ \\ 
Figure 7. Contour plots voltage potential near the upper quantum dot of the qubit: (a) when the state of  qubit is \br{1}; (b) when state is \br{0}. A  shows the measuring point.


\newpage
\psfig{file=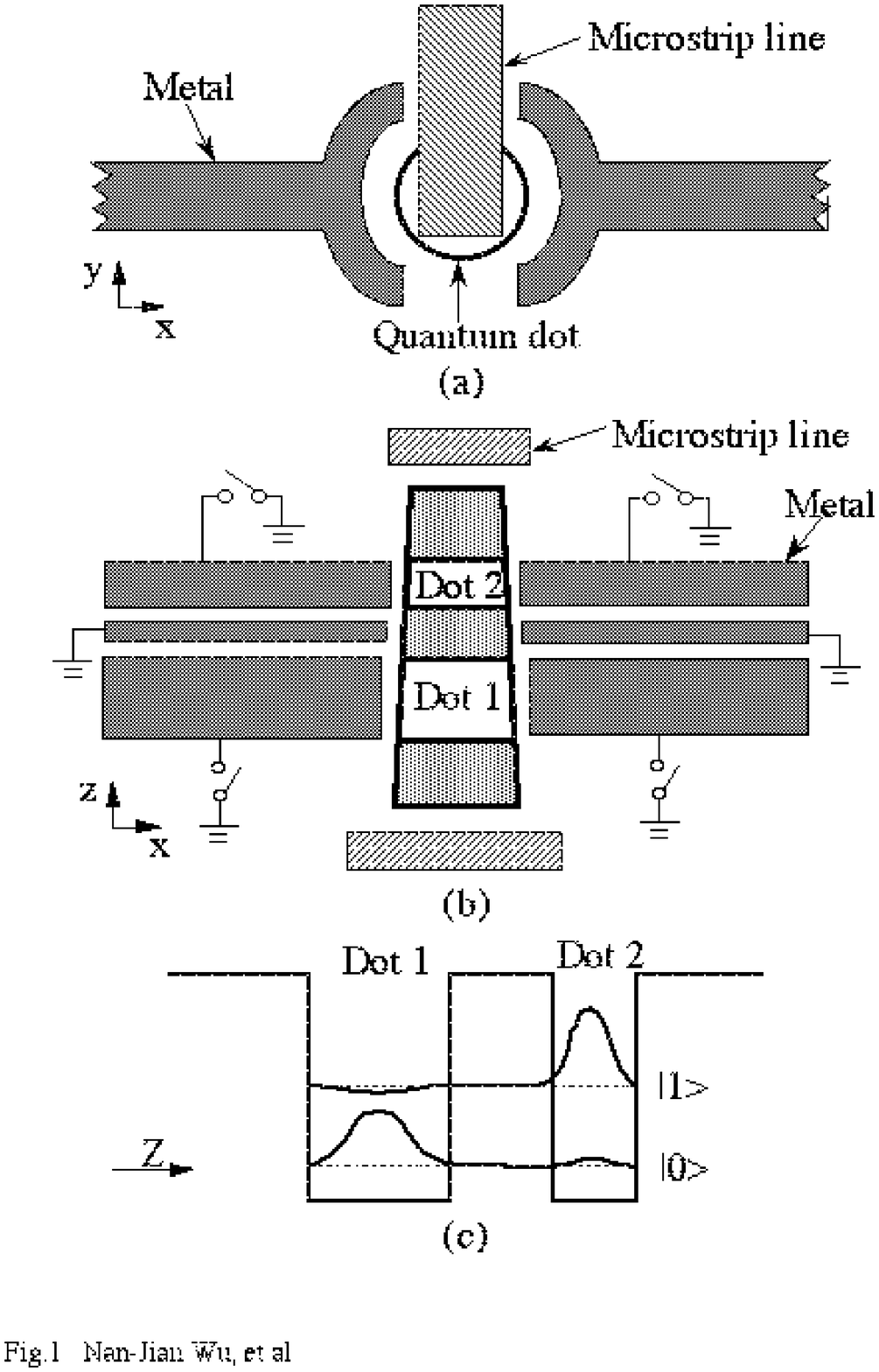}
\newpage
\psfig{file=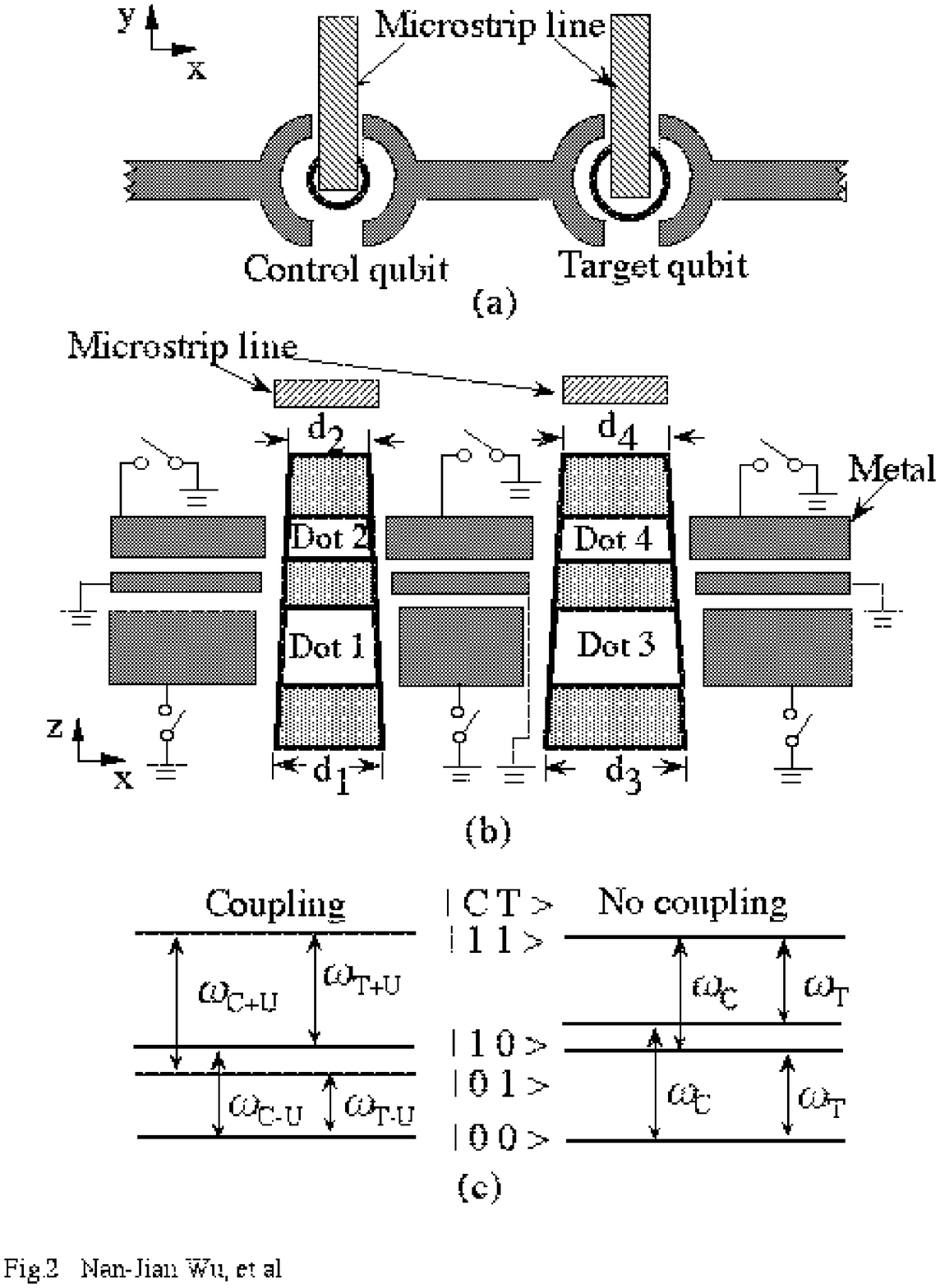}
\newpage
\psfig{file=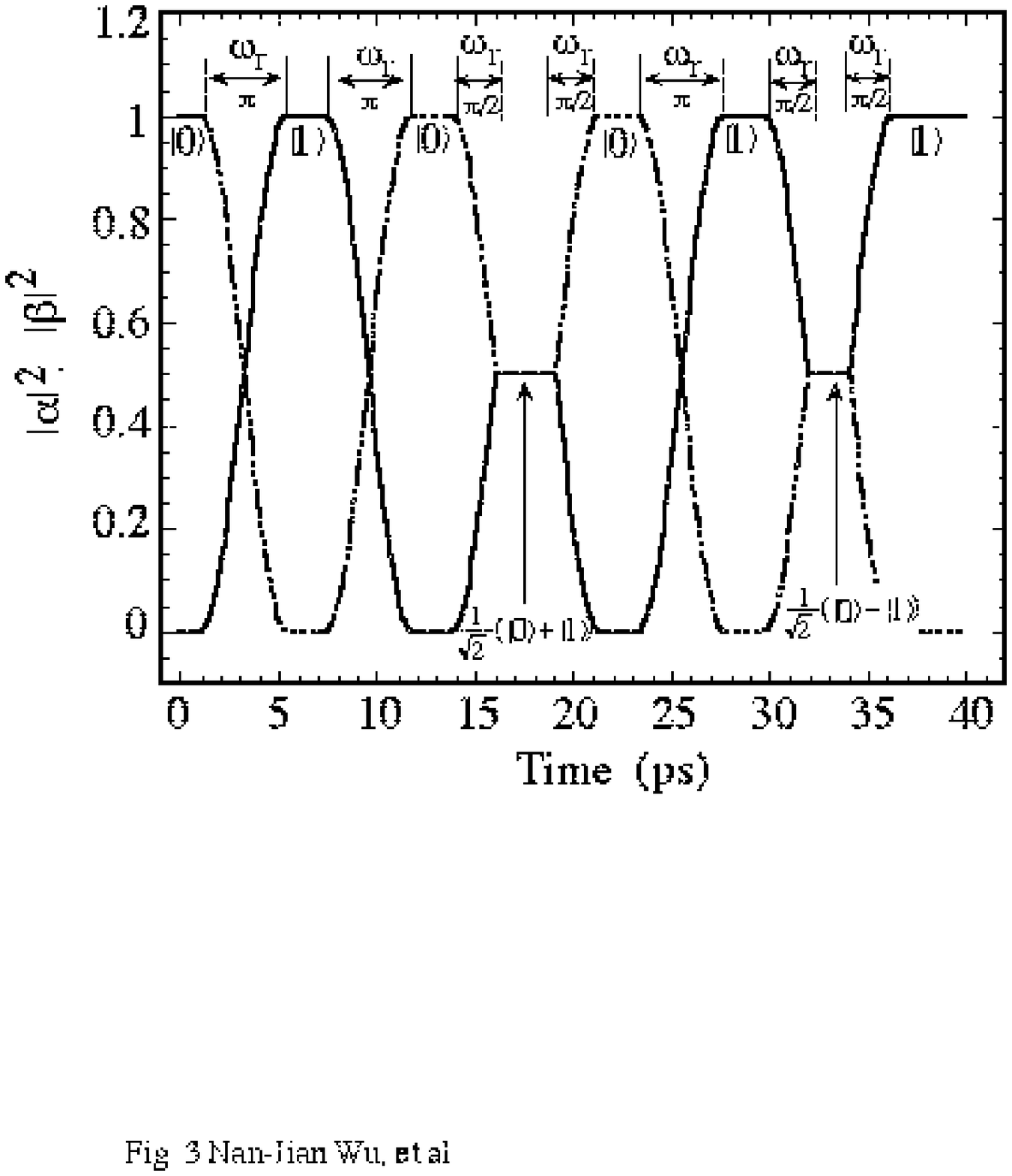}
\newpage
\psfig{file=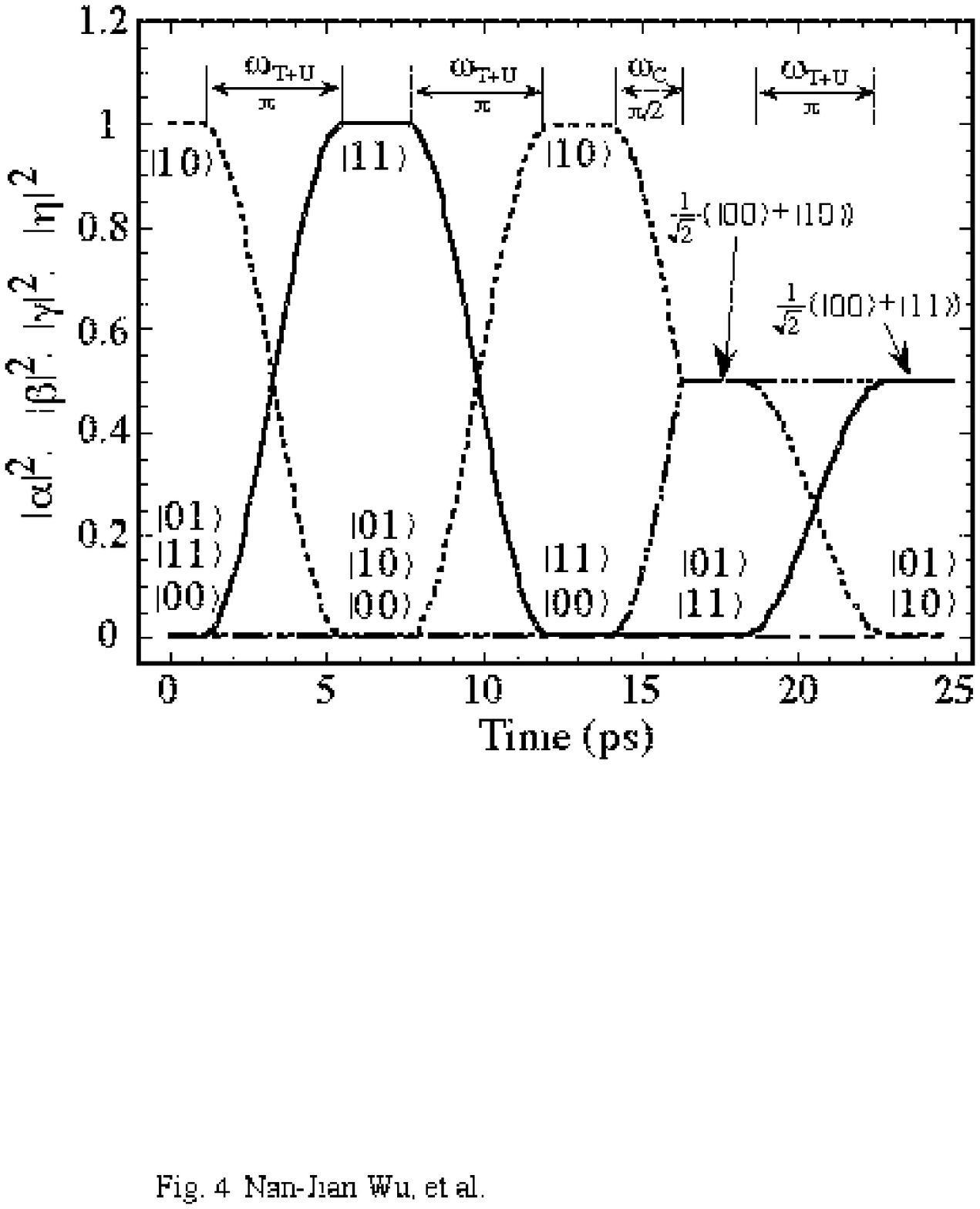}
\newpage
\psfig{file=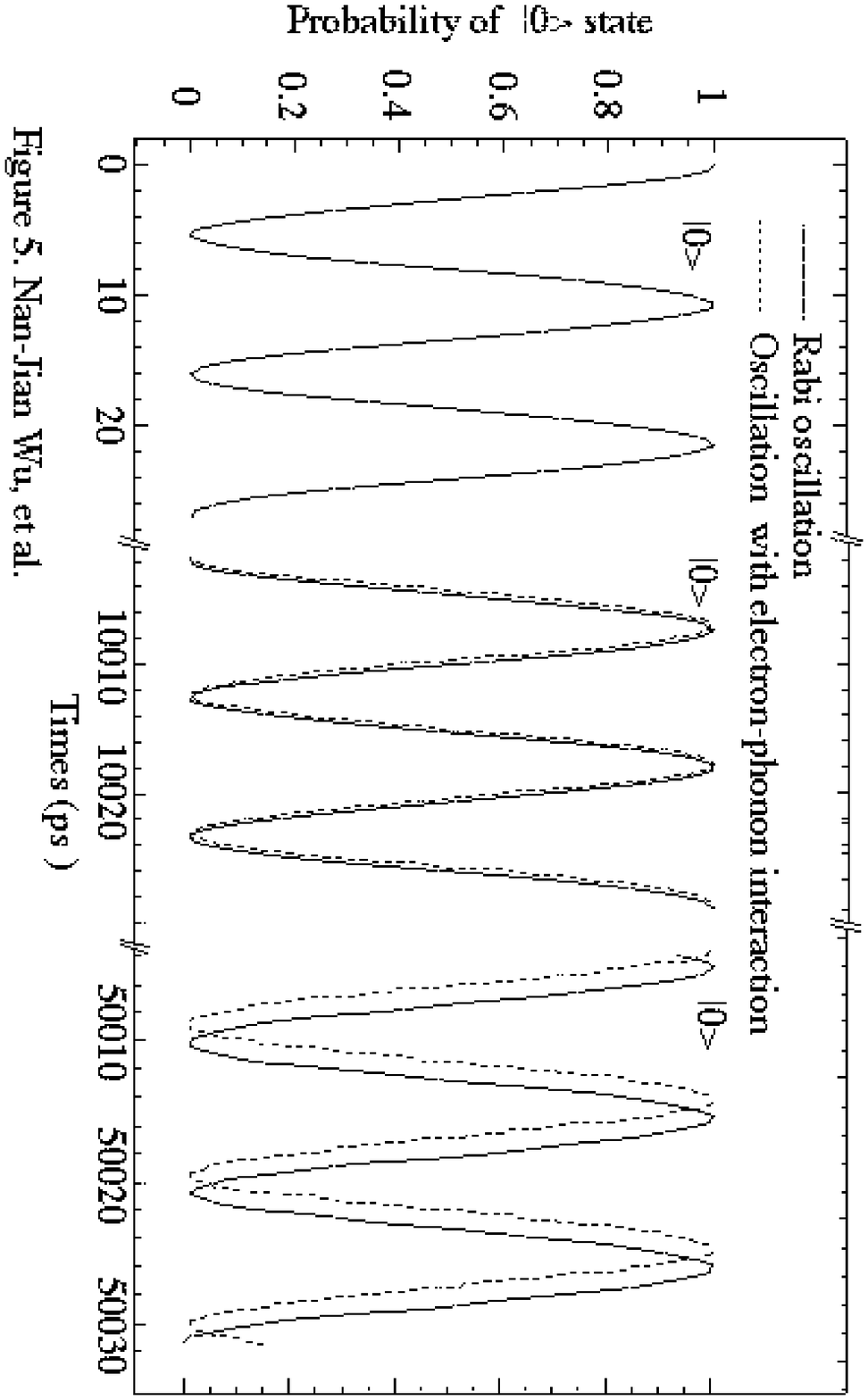}
\newpage
\psfig{file=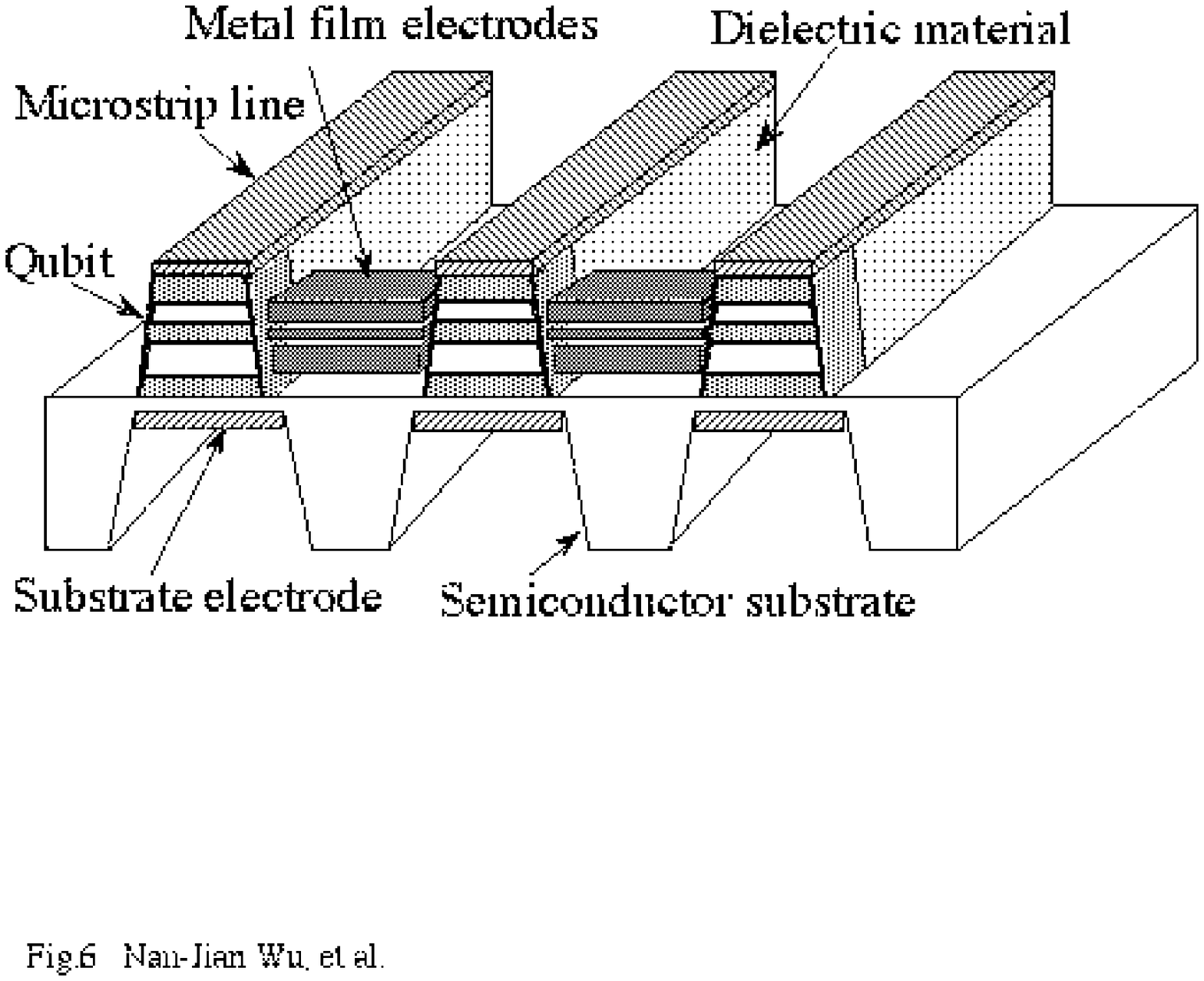,width=17cm}
\newpage
\psfig{file=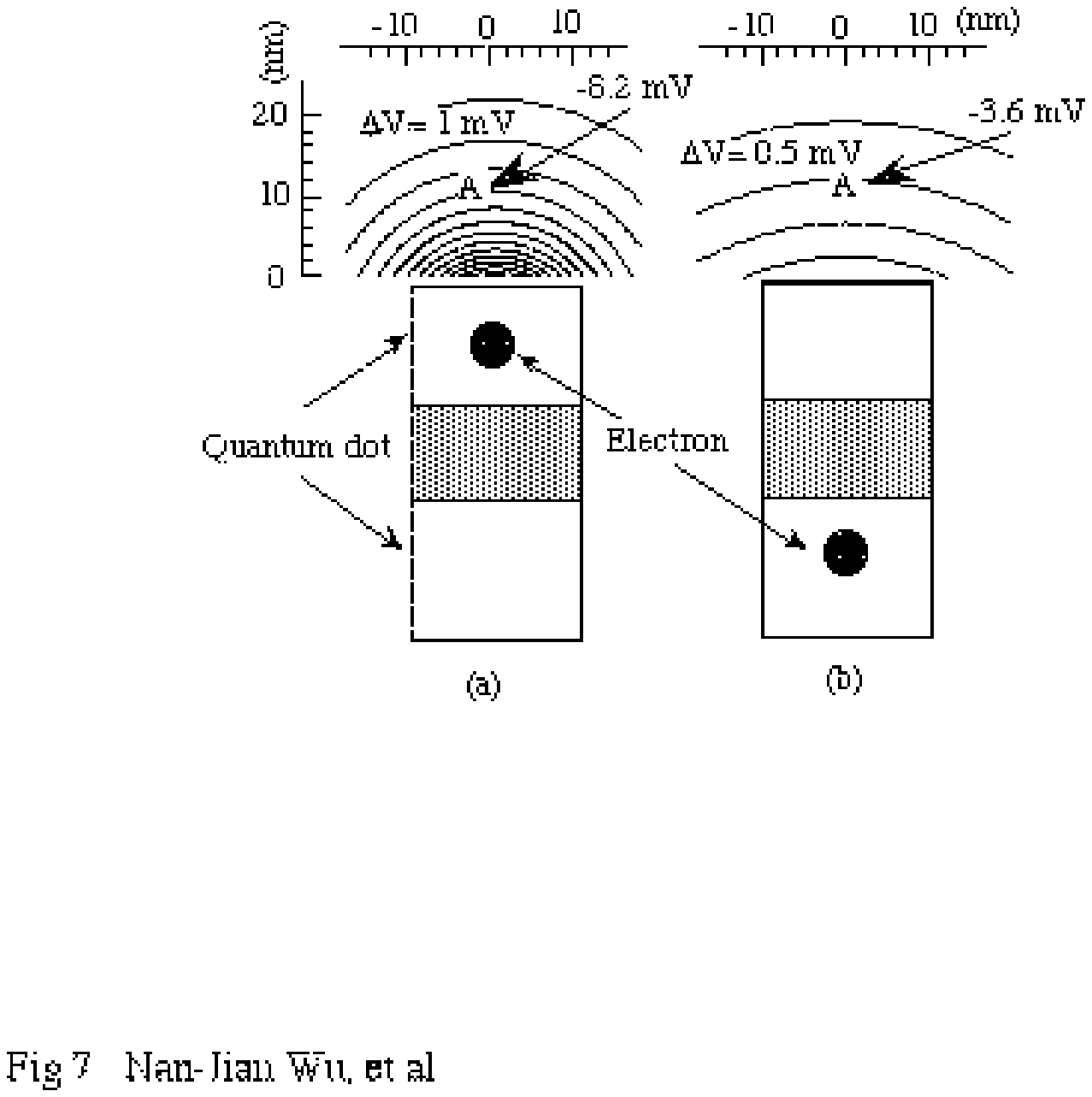}


\begin{thebibliography}{9}
\bibitem{Barenco}
A. Barenco, D. Deutsch, A. Ekert and R. Jozsa, Phys. Rev. Lett. {\bf 74}(1995)4083.
\bibitem{}
D. Loss and D. P. DiVincenzo, Phys. Rev. A {\bf 57}(1998) 120.
\bibitem{}
B. E. Kane, Nature, {\bf 393} (1998) 133.
\bibitem{}
A. Shnirman, G. Schon and Z. Hermon, Phys. Rev. Lett. {\bf 79}(1997)2371.
\bibitem{}
Y. Nakamura, Y. Pashkin and J. S. Tsai, Nature, {\bf 398}(1999)786.
\bibitem{}
T. Schmidt, R. J. Hang, K. V. Klizing, A. Forster and H. Luth, Phys. Rev. Lett. {\bf 78}(1997)1544. 
\bibitem{}
D. G. Austing, T. Honda, K. Muraki, Y. Tokura, S. Tarucha, Physica B {\bf 249-252} (1998) 206.
\bibitem{}
J. H. Davies, The Physics of Low-Dimensional Semiconductor, (Cambridge University Press,1993)
\bibitem{}
W. Wu, J. R. Tucker, G. S. Solomon and J. S. Harris, Appl. Phys. Lett., {\bf 71}(1997)1083. 
\bibitem{}
 T. Fujisawa, T. H. Oosterkamp, W. G. van der Wiel, B. W . Broer, R. Aguado, S. Tarucha, and L. P. Kouwenhoven, Science {\bf 282}(1998)932 .
\bibitem{}
K. C. Gupta, et al, Microstrip Lines and Slotlines, second Edition, (Artech House, Boston and London, 1996)
\bibitem{}
K. Nomoto, R. Ugajin, T. Suzuki and I. Hase, J. Appl. Phys., {\bf 79}(1996)291. 

\end{thebibliography}
\end{document}